# Uncertainty relation and quantitative condition of inversion symmetry of Time in many-particle system

You-gang Feng

Department of Basic Science, College of Science, Guizhou University,

Cai Jia Guan, Guiyang 550003, China

## Abstract

We proved that the uncertainty relation fits in with many-particle system and the equality of the relation corresponds to the thermodynamic equilibrium state, the inequality of the relation corresponds to the thermodynamic non-equilibrium state for any quantum system. The microscopic origin of the second law of thermodynamics is certainly resulted in the wave-particle duality of matter. A quantitative condition of inversion symmetry of time is obtained.

*PACS(2003): 05.30.-d, 05.70.-a, 03.65.-w*

## . Introduction

We proved [1] that the equality of the uncertainty relation corresponds to thermodynamic equilibrium state, the inequality does to the thermodynamic non-equilibrium state; the microscopic origin of the second law of thermodynamics is the wave-particle duality of matter. In our proving the wave functions, which describe the systems, are single-particle wave functions. If the wave functions are many-particle wave functions, from which we will not able to derive the uncertainty relation [2]. The problem is whether the uncertainty relation satisfies to the many-particle system ? If many-particle system can be described by single-particle wave functions we can link the uncertainty relation with the system because the uncertainty relation can be proved by the single-particle wave function.

In fact, the states of many-particle system can be expressed by the single-particle wave functions solved by means of Hartree-Fock-Dirac self-consistent field method [3]. However, the method is of approximate solution. Whether there exists an exact method to describe the many-particle system by the single-particle wave functions? We do not need to solve practically the system's equation, we only need to prove theoretically that a many-particle system can be exactly described by the single-particle wave functions. If it is true the uncertainty relation will fit in with the many-particle system. In the following discussion we neglect electron spins as they do not affect the proving of the relation.

. Theory

For one-dimension let a molecule be made up of m nuclei and s electrons, and a many-particle system consist of a large number of these molecules. The distances between nuclei are fixed relatively in a steady state because each nuclear' mass is by far greater than the electrons' masses, and we only need to consider the position coordinates of s electrons. In the coordinate representation the wave function of a molecule with s classical freedoms is denoted as $|\Psi(q_1, q_2, ..., q_s, t)\rangle$, where $q_1, ..., q_s$ are s general coordinates corresponding to the s classical freedoms, $t$ labels time. We noticed that according to the quantum statistical mechanics [4], a series of wave functions in the probability-density operator of the many-particle system are the wave functions at different times, this fact gives our reason for describing the system by the single-particle wave functions. Let the position coordinate of the $i-th$ electron be $q_{i1} = x_{i1}$ at time $t_{i1}$, the position coordinates of other (s-1) electrons be taken by a real-constant set $Q_1 = (q_1^1, ..., q_{i-1}^1, q_{i+1}^1, ..., q_s^1)$; the position coordinate of the $i-th$ electron be $q_{i1}' = x_{i1}' \neq x_{i1}$ at time $t_{i1}'$, and the position coordinates of other (s-1) electrons be written by the same set $Q_1$,..., and denote the state of the $i-th$ electron corresponding to the set $Q_1$ as a wave function $|\psi^{i1}\rangle$, which exists certainly due to Ergodic hypthesis. Obviously, $|\psi^{i1}\rangle$ is a single-particle wave function. When the position coordinates of other (s-1) electrons change from the set $Q_1$ to a new real-constant set $Q_2 = (q_1^2, ..., q_{i-1}^2, q_{i+1}^2, ..., q_s^2)$, $Q_1 \neq Q_2$, the state of the $i-th$ electron is denoted as $|\psi^{i2}\rangle$, ..., similarly, the state $|\psi^{ij}\rangle$ corresponds to a real-constant set $Q_j = (q_1^j,...,q_{i-1}^j,q_{i+1}^j,...,q_s^j)$. Generally, $Q_p \neq Q_j$ if $p \neq j$, and, $n \to +\infty$. Thus, the wave functions of the $i-th$ electron are a series of functions: $|\psi^{i1}\rangle, |\psi^{i2}\rangle,...,|\psi^{ij}\rangle,...,|\psi^{ip}\rangle,...,|\psi^{in}\rangle$. Generally, $Q_p \neq Q_j$ if $p \neq j$, and, $n \to +\infty$. Each of them relates to a special state of the many-particle system at a certain time. Let $N_{ij}$ subsystems be in state $|\psi^{ij}\rangle$ in $N_i$ subsystems, which corresponds to the real-constant set $Q_j = (q_1^j,...,q_{i-1}^j,q_{i+1}^j,...,q_s^j)$ and $N_i = \sum_j N_{ij}$. Therefore, a mixed ensemble consists of $N$ subsystems, $N = \sum_i N_i = \sum_i \sum_j N_{ij}$, so that the probability-density operator of the ensemble is defined as

$$\hat{\rho} = \sum_i \sum_j (N_{ij}/N) |\psi^{ij}\rangle\langle\psi^{ij}| = \sum_i \rho_i \sum_j |\psi^{ij}\rangle\langle\psi^{ij}| \qquad (1)$$

where

$$\rho_{ij} = N_{ij}/N_i \;,\quad \rho_i = N_i/N \;,\quad \rho_i \rho_{ij} = N_{ij}/N \qquad (2.a)$$

$$\sum_i \rho_i = 1 \quad,\quad \sum_j \rho_{ij} = 1 \qquad (2.b)$$

Since $|\psi^{ij}\rangle$ is a single-particle wave function the uncertainty relation fits for the many-particle system, which is given by

$$(\Delta x_{ij})^2 (\Delta p_{x_{ij}})^2 \geq (h/4\pi)^2 \qquad (3)$$

where

$$(\Delta x_{ij})^2 = \langle(x - \langle x_{ij}\rangle)^2\rangle = \langle\psi^{ij}|(x - \langle x_{ij}\rangle)^2|\psi^{ij}\rangle \qquad (4.a)$$

$$(\Delta p_{x_{ij}})^2 = \langle(p_x - \langle p_{x_{ij}}\rangle)^2\rangle = \langle\psi^{ij}|(p_x - \langle p_{x_{ij}}\rangle)^2|\psi^{ij}\rangle \qquad (4.b)$$

$$\langle x_{ij}\rangle = \langle\psi^{ij}|x|\psi^{ij}\rangle \qquad (4.c)$$

Using of Eqs.(2.b) and (3), we get

$$(\Delta x_i)^2 (\Delta p_{x_i})^2 = \sum_j \rho_{ij} (\Delta x_{ij})^2 (\Delta p_{x_{ij}})^2 \geq \sum_j \rho_{ij} (h/4\pi)^2 = (h/4\pi)^2$$

namely, there is an uncertainty relation in $N_i$ subsystems:

$$(\Delta x_i)^2 (\Delta p_{x_i})^2 \geq (h/4\pi)^2 \qquad (5)$$

Eq.(5) indicates that a many-particle system represented by state $|\psi^{ij}\rangle$ obeys the uncertainty relation. With the same reason as used in reference [1] in the $N_i$ subsystems the probability-density function of random fluctuations in the thermodynamic equilibrium state is written by

$$f_i(x, p_x) = \frac{1}{2\pi(\Delta x_i)(\Delta p_{x_i})} \exp\{-\frac{1}{2}[\frac{(x-\langle x_i \rangle)^2}{(\Delta x_i)^2} + \frac{(p_x - \langle p_{x_i} \rangle)^2}{(\Delta p_{x_i})^2}]\} \qquad (6)$$

Eq.(6) is similar to Eq.(4) of reference [1], however, here the number $i$ labels the $i-th$ electron, and $i = 1,2,...,s$.

For the same reason we can treat other (s-1) electrons. So far we can say the whole many-particle system can be described by the single-particle wave functions of s electrons at different times. For the ensemble, we have

$$\langle x \rangle = Tr(\hat{\rho} x) = \sum_i \rho_i \langle x_i \rangle \qquad (7.a)$$

$$\langle p_x \rangle = Tr(\hat{\rho} p_x) = \sum_i \rho_i \langle p_{x_i} \rangle \qquad (7.b)$$

so that

$$(\Delta x)^2 (\Delta p_x)^2 = \sum_i \rho_i (\Delta x_i)^2 (\Delta p_{x_i})^2 \geq \sum_i \rho_i (h/4\pi)^2$$

Using of Eq.(2.b), we obtain

$$(\Delta x)^2 (\Delta p_x)^2 \geq (h/4\pi)^2 \qquad (8)$$

Eq.(8) means that the uncertainty relation is still hold for a many-particle system. In fact, in a many-body molecular in a steady state $\langle x \rangle$ is the position of its negative charges' center which has a certain meaning, and the position coordinates of each electron of the molecule fluctuate about $\langle x \rangle$.

The probability-density function of random fluctuations of the ensemble is as the same form as given by Eq.(7) of reference [1], in brief, which we omit here.

## . Discussion

A limitation of the second law of thermodynamics is inversion symmetry of time which is a reversible process, for examples, a zygote grew into a baby girl after her twelve-years' freezing, a lotus seed buried underground over one thousand years sprouted up again,…, all these, as they underwent a special period—inversion symmetry of time. It seems that time lost their directional arrow, all things stopped their developing during the period. What is its quantitative condition? If we understand and control this law, we will prolong human life, freshen foods and preserve historical objects more efficiently. As the time is inversion symmetry in the equilibrium state the minimal uncertainty relation gives a fixed quantitative condition of the inversion symmetry of time [1]:

$$\Delta t = h/(4\pi \Delta E) \qquad (9)$$

Here we must emphasize that an observer in the equilibrium state of the system is not able to feel the existence of the $\Delta t$, because the time is inverse and he can not be aware of any interval of the time. There are not "before" and "future", but only "now" for him, he himself can not write the Eq.(9) which is written by the observers who are in the time-ordered system and out of the thermodynamic equilibrium state of the system. In a view of the observers who are in the time-ordered system the equilibrium state will exist for the interval of time $\Delta t$.

. Conclusion

Finally, we make conclusion that for any quantum system when random variable fluctuations obey the central limit theorem the equality of the uncertainty relation corresponds to the thermodynamic equilibrium state. The inequality corresponds to the thermodynamic non-equilibrium state. The uncertainty relation is a quantum mechanics expression of the second law of thermodynamics originated in the wave-particle duality of matter. The fluctuation of the system's energy controls the period of inversion symmetry of time, which has great meaning for genetic engineering projects, medical treatment, life science, food industry, archaeology and preserving of cultural objects and historical monuments, etc.